\documentclass[prb,twocolumn,superscriptaddress,showpacs,epsf]{revtex4}

\usepackage{graphicx}
\usepackage{latexsym}
\usepackage{amssymb}
\usepackage{amsfonts}
\usepackage{soul}
\usepackage{color}
\begin{document}
\bibliographystyle{apsrev}

\title{Mesoscopic cross--film cryotrons:
Vortex trapping and \\ dc--Josephson--like oscillations of the
critical current}

\author{A.Yu. Aladyshkin}
\affiliation{INPAC -- Institute for Nanoscale Physics and
Chemistry, K.U. Leuven, Celestijnenlaan 200D, B--3001 Leuven,
Belgium} \affiliation{Institute for Physics of Microstructures
RAS, 603950, Nizhny Novgorod, GSP-105, Russia}
\author{G.W.
Ataklti} \affiliation{INPAC -- Institute for Nanoscale Physics and
Chemistry, K.U. Leuven, Celestijnenlaan 200D, B--3001 Leuven,
Belgium}
\author{W. Gillijns}
\affiliation{INPAC -- Institute for Nanoscale Physics and
Chemistry, K.U. Leuven, Celestijnenlaan 200D, B--3001 Leuven,
Belgium}
\author{I.M. Nefedov}
\affiliation{Institute for Physics of Microstructures RAS, 603950,
Nizhny Novgorod, GSP-105, Russia}
\author{I.A. Shereshevsky}
\affiliation{Institute for Physics of Microstructures RAS, 603950,
Nizhny Novgorod, GSP-105, Russia}
\author{A.V. Silhanek}
\affiliation{INPAC -- Institute for Nanoscale Physics and
Chemistry, K.U. Leuven, Celestijnenlaan 200D, B--3001 Leuven,
Belgium}
\author{J. Van de Vondel}
\affiliation{INPAC -- Institute for Nanoscale Physics and
Chemistry, K.U. Leuven, Celestijnenlaan 200D, B--3001 Leuven,
Belgium}
\author{M. Kemmler}
\affiliation{Physikalisches Institut -- Experimentalphysik II and
Center for Collective Quantum Phenomena, Universit\"{a}t
T\"{u}bingen, Auf der Morgenstelle 14, 72076 T\"{u}bingen,
Germany}
\author{R. Kleiner}
\affiliation{Physikalisches Institut -- Experimentalphysik II and
Center for Collective Quantum Phenomena, Universit\"{a}t
T\"{u}bingen, Auf der Morgenstelle 14, 72076 T\"{u}bingen,
Germany}
\author{D. Koelle}
\affiliation{Physikalisches Institut -- Experimentalphysik II and
Center for Collective Quantum Phenomena, Universit\"{a}t
T\"{u}bingen, Auf der Morgenstelle 14, 72076 T\"{u}bingen,
Germany}
\author{V.V. Moshchalkov}
\affiliation{INPAC -- Institute for Nanoscale Physics and
Chemistry, K.U. Leuven, Celestijnenlaan 200D, B--3001 Leuven,
Belgium}

\date{\today}
\begin{abstract}
We investigate theoretically and experimentally the transport
properties of a plain Al superconducting strip in the presence of
a single straight current-carrying wire, oriented perpendicular to
the superconducting strip. It is well known that the critical
current of the superconducting strip, $I_c$, in such
cryotron--like system can be tuned by changing the current in the
control wire, $I_w$. We demonstrated that the discrete change in
the number of the pinned vortices/antivortices inside the narrow
and long strip nearby the current-carrying wire results in a
peculiar oscillatory dependence of $I_c$ on $I_w$.
\end{abstract}




\maketitle

\section{Introduction}

The original idea to control the resistance of a long
superconducting (S) wire by means of a magnetic field generated by
a coil wound locally over the wire, was proposed by Buck in
1956.\cite{Buck-56} If the magnetic field inside the solenoid with
permanent driving current exceeds the critical field of the
central type--I superconducting wire, superconductivity will be
completely suppressed and this device (cryotron) will be switched
from a low-resistive state to a high-resistive state. Further
investigations in the 1960's
\cite{deGennes-66,Bremer-62,Lock-62,Newhouse-69} revealed that the
cryotrons can be potentially used as superconducting computer
elements (switches, binary adders, shift registers, AND/OR gates
etc). However the performance of such elements at high frequencies
was found to be worse than ordinary semiconducting chips, since
the cryotrons were rather slow and energy-consuming. Furthermore,
the miniaturization of cryotron--like devices, which appears to be
important for reducing the characteristic time constant, was very
limited at those times.

Revival of interest to superconductor--electromagnet (S/Em)
hybrids came in the 1990's in connection with the problem of the
influence of a spatially-modulated magnetic field on
superconductivity. The development of advanced techniques for
material deposition and lithographic methods have made it possible
to fabricate composite structures with controlled arrangement of
superconducting, normal metallic, ferromagnetic and insulating
layers at micron and submicron levels. Pannetier {\it et al.}
\cite{Pannetier-95} demonstrated that the dependence of the
critical temperature $T_c$ on the applied magnetic field $H$ for
such a hybrid system consisting of a plain Al film and a
lithographically defined array of parallel metallic lines can be
nonlinear and even non-monotonous in contrast to a plain
superconducting film in a uniform magnetic field.\cite{Tinkham}
Such tunability of the $T_c(H)$ dependence by the stray field of
the meander--like wire reflects directly the controllable
modification of the standard Landau spectrum for the order
parameter (OP) wave function in a periodic magnetic
field.\cite{Peeters,Aladyshkin-PRB-03,Aladyshkin-PRB-06} It is
worth noting that keeping the nonuniform magnetic field of the
control coils/wires requires expenditure of energy and may result
in parasitic heating effects. Therefore, later on the interest was
turned to superconductor--ferromagnet (S/F) hybrids since the
inhomogeneous magnetic field in S/F systems, conditioned by the
nonuniform distribution of magnetization, can be obtained without
energy costs. The influence of a nonuniform magnetic field
produced by ferromagnetic elements on nucleation of
superconductivity and low-temperature properties of
superconducting specimens was studied intensively for flux-coupled
S/F hybrids during the last two decades (see reviews
\cite{Lyuksyutov-AdvPhys-05,Velez-JMMM-08,Aladyshkin-SuST-09} and
references therein). We would like to note that the amplitude of
the stray magnetic field and its profile are dictated by the
saturated magnetization of the ferromagnet and by the shape of the
ferromagnetic elements, which eventually restricts the flexibility
of the S/F hybrids.

In the present paper we report on a so far unexplored limit of a
cryotron--like system at micro- and nanoscales bringing together
the ideas and approaches developed for both S/Em and S/F hybrids.
With that purpose we fabricated a composite structure consisting
of a {\it type-II} superconducting strip and a single straight
current-carrying S wire oriented perpendicular to the strip. Since
the stray magnetic field of the straight wire is maximal near the
wire and it decays approximately as $1/r$ at large distances $r$
from the wire, one can expect that only a small part of the
superconducting strip in close vicinity to the wire will be
affected by the nonuniform magnetic field.\cite{Sonin} A simple
change in the current in the wire, $I_w$, gives us the opportunity
to control the maximal value of the magnetic field, $B_0$,
generated by the current in the control wire. By varying $B_0$ one
can create different states with {\em partially} and completely
suppressed superconductivity in the considered cross--film
cryotron. Thus, the main point of our research is to ``scan" all
possible intermediate states lying between uniform
superconductivity (for small $B_0$) and fully depleted
superconductivity (for large $B_0$) and to study the effect of the
transitions between various superconducting states on the critical
current for mesoscopic type--II cross--film cryotrons.

    \begin{figure}[!b]
    \includegraphics[width=6.5cm]{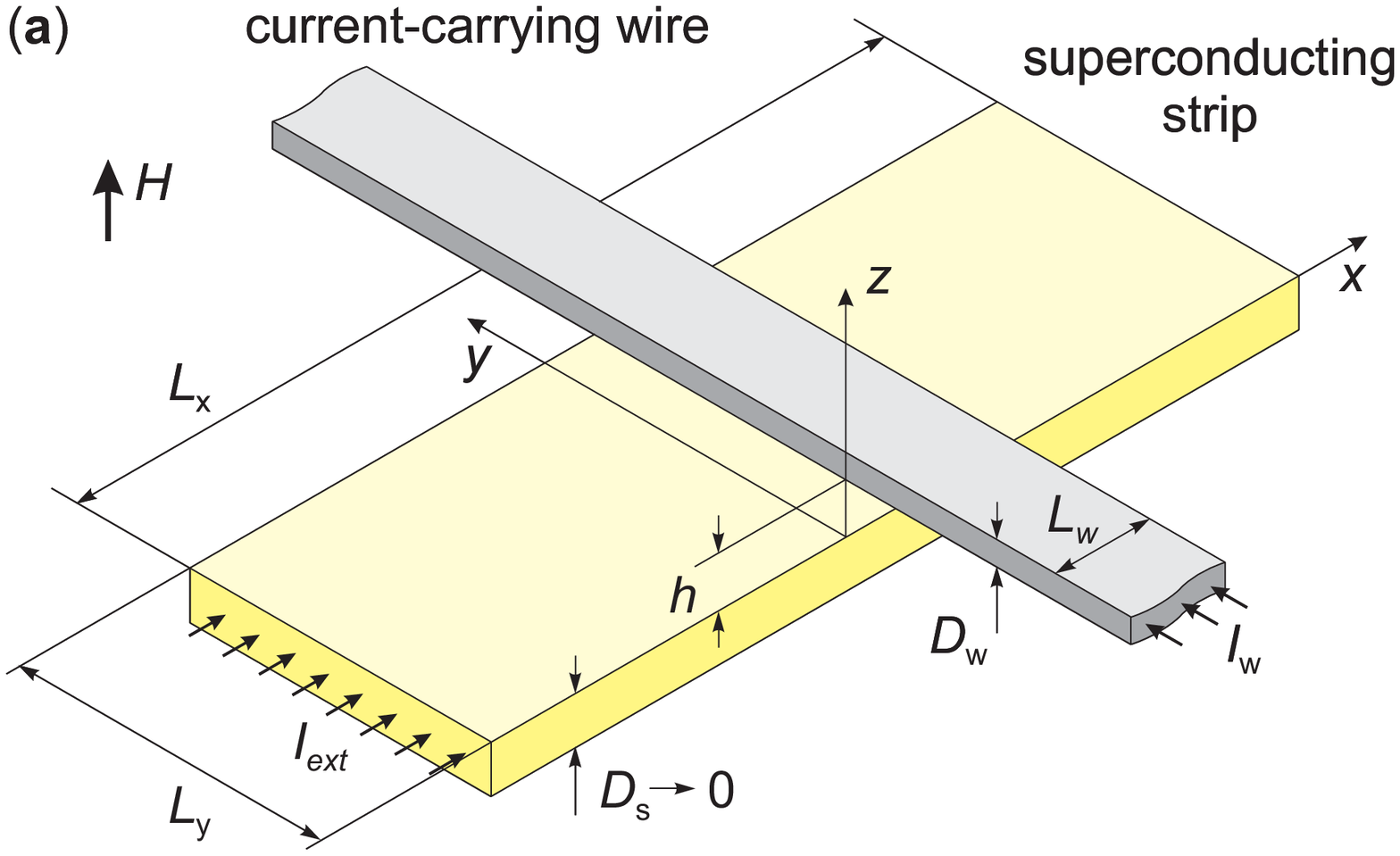}\\
    \includegraphics[width=6.5cm]{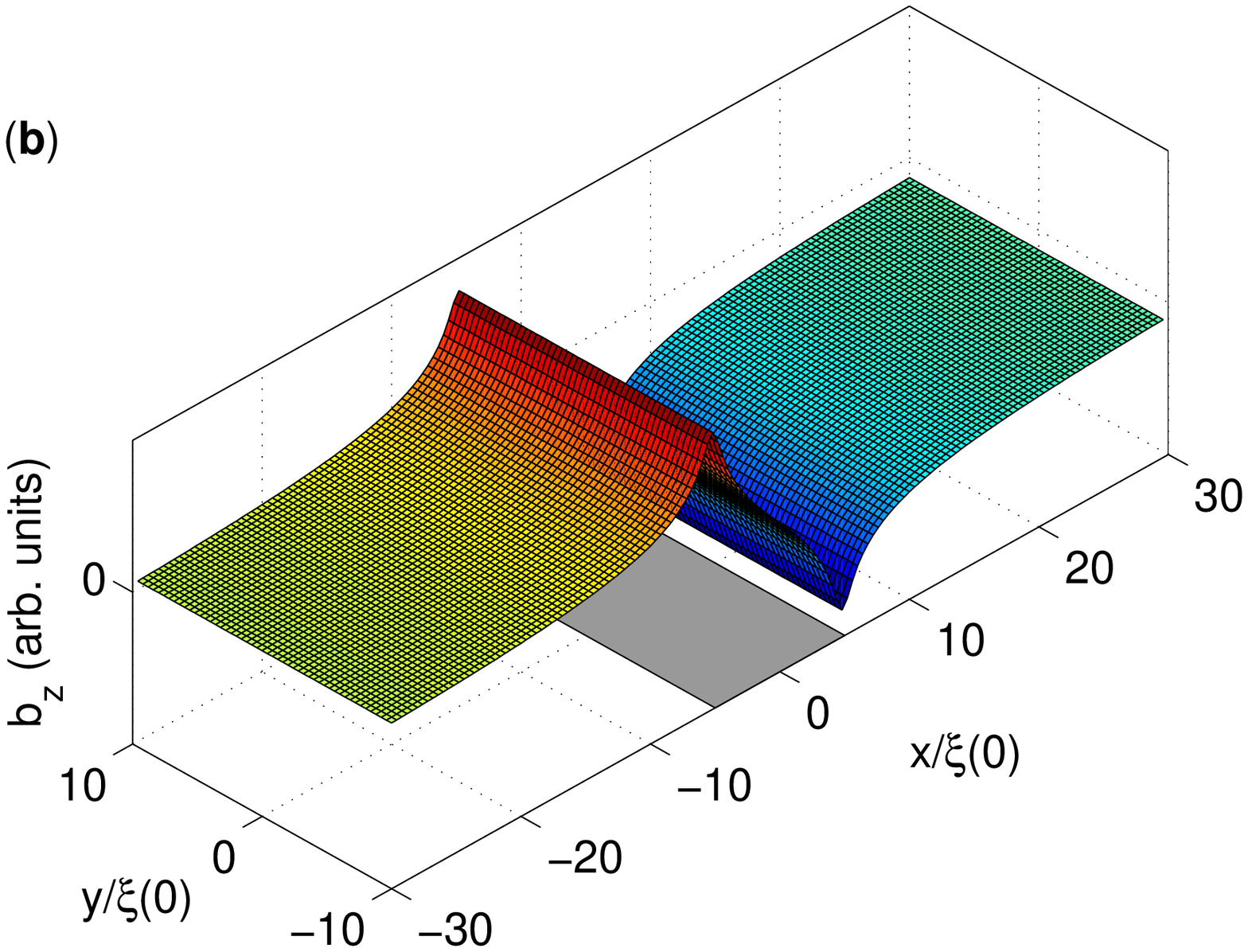}\\
    \caption{(color online) (a) Schematic presentation of the mesoscopic cross--film cryotron.
    (b) Profile of the $z-$component of the magnetic field, induced
    by the current-carrying wire, in the middle plane of the superconducting strip.
    The maximal and minimal $b_z$ values correspond to the edges of the
    current-carrying wire (shown in grey). \label{Fig-System}}
    \end{figure}

\section{Theory}

\subsection{Model}

To simulate the {\it onset} of vortex dynamics in the
superconducting strip in the field of the current-carrying wire
(i.e. in the cross--film cryotron) under the action of the
transport current we use a time--dependent Ginzburg--Landau
model.\cite{IvlevKopnin-84} The thickness $D_s$ of the
superconducting strip is assumed to be infinitely small, so the
effective magnetic field penetration length
$\Lambda=\lambda_L^2/D_s$ exceeds substantially the lateral
dimensions of the superconducting sample ($\lambda_L$ is the
London penetration depth). Consequently, both the magnetic field
${\bf B}={\rm rot\,}{\bf A}$ and the vector potential ${\bf A}$
are determined solely by external sources. We use the following
units: $m^*\sigma_n \beta/(2e^2\tilde{\alpha})$ for time, the
coherence length $\xi(0)$ at temperature $T=0$ for distances,
$\Phi_0/(2\pi\xi(0))$ for the vector potential and
$4e\tilde{\alpha}^2\xi(0)/(\hbar \beta)$ for the current density,
where $\alpha=-\tilde{\alpha}\,\tau$ and $\beta$ are the
conventional parameters of the Ginzburg-Landau expansion,
$\tau=(1-T/T_{c0})$, $T_{c0}$ is the critical temperature at
$B=0$, $e$ and $m^*$ are charge and the effective mass of
carriers, $\sigma_n$ is the normal state conductivity. Then the
time--dependent Ginzburg--Landau equations take the form
\cite{Kapra-11}

    \begin{eqnarray}
    u \left(\frac{\partial}{\partial t}+ i\varphi
    \right) \psi = \tau\,\left(\psi-|\psi|^2\psi\right)
    + \left(\nabla + i{\bf A}\right)^2\psi, \\
    \nabla^2 \varphi = {\rm div}\, {\bf j}_s, \quad
    {\bf j}_s = -\frac{i}{2} \tau\,
    \Big\{\psi^*\left(\nabla + i{\bf A}\right)\psi -
    \mbox{c.c.}\Big\}.
    \end{eqnarray}
Here $\psi$ is the normalized order parameter (OP), $\varphi$ is
the dimensionless electrical potential, ${\bf j}_s$ is the
dimensionless density of superconducting currents, $u$ is the rate
of the OP relaxation, c.c. stands for complex conjugate. To
complete the problem we apply the boundary conditions
    \begin{eqnarray}
    \left(\frac{\partial}{\partial n} + i A_n\right)_{\Gamma} \psi =
    0,\quad \left(\frac{\partial \varphi}{\partial n} \right)_{\Gamma} =
    j_{ext},
    \label{BoundCond-1}
    \end{eqnarray}
which seems to be natural for the superconductor/vacuum or
superconductor/insulator interfaces, $n$ is the normal vector to
the sample's boundary $\Gamma$, $j_{ext}$ is the normal component
of the inward (outward) flow of the external current. To detect a
vortex on the grid cell corresponding to the certain grid node
($x_i,y_j$), we use the following criterion:\cite{Melnikov-PRB-02}
    \begin{eqnarray}
    \nonumber \arg (\psi^*_{i,j}\psi_{i,j+1}) + \arg (\psi^*_{i,j+1}\psi_{i+1,j+1})
    \\ + \arg (\psi^*_{i+1,j+1}\psi_{i+1,j}) + \arg (\psi^*_{i+1,j}\psi_{i,j}) = 2\pi N,
    \end{eqnarray}
$\psi_{i,j}=\psi(x_i,y_j)$, $\arg(\psi)$ is the argument of the
complex number $\psi_{i,j}$, belonging to the interval (-$\pi$,
$\pi$]. If $N=+1$, there is a vortex in the cell, while $N=-1$
corresponds to an antivortex.  It is important to note that we
define the vortex (antivortex) as a position of the singularity of
the OP phase, $\theta = \arg(\psi)$, however the direction, in
which the superconducting currents circulate around the center of
the vortex (antivortex), depends on the charge of carriers (i.e.
the $e$ sign).

    \begin{figure}[t!]
    \begin{center}
    \includegraphics[width=6.9cm]{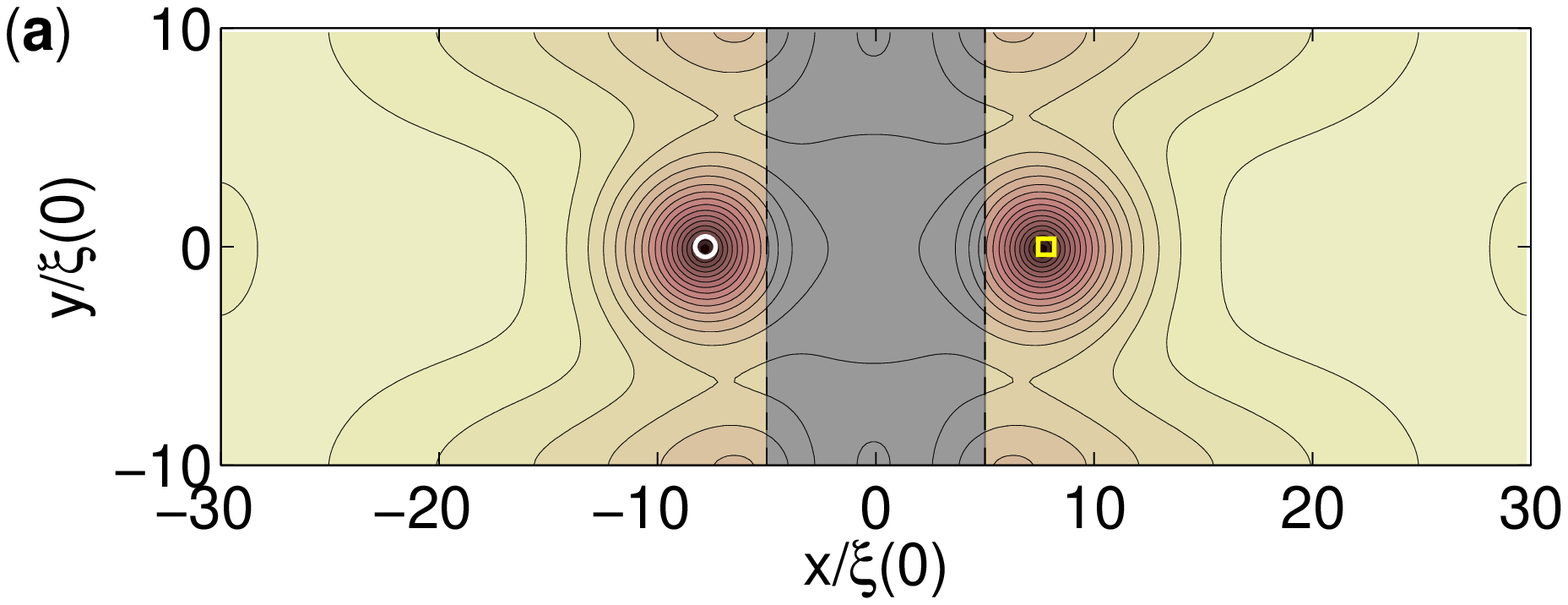}
    \includegraphics[width=6.9cm]{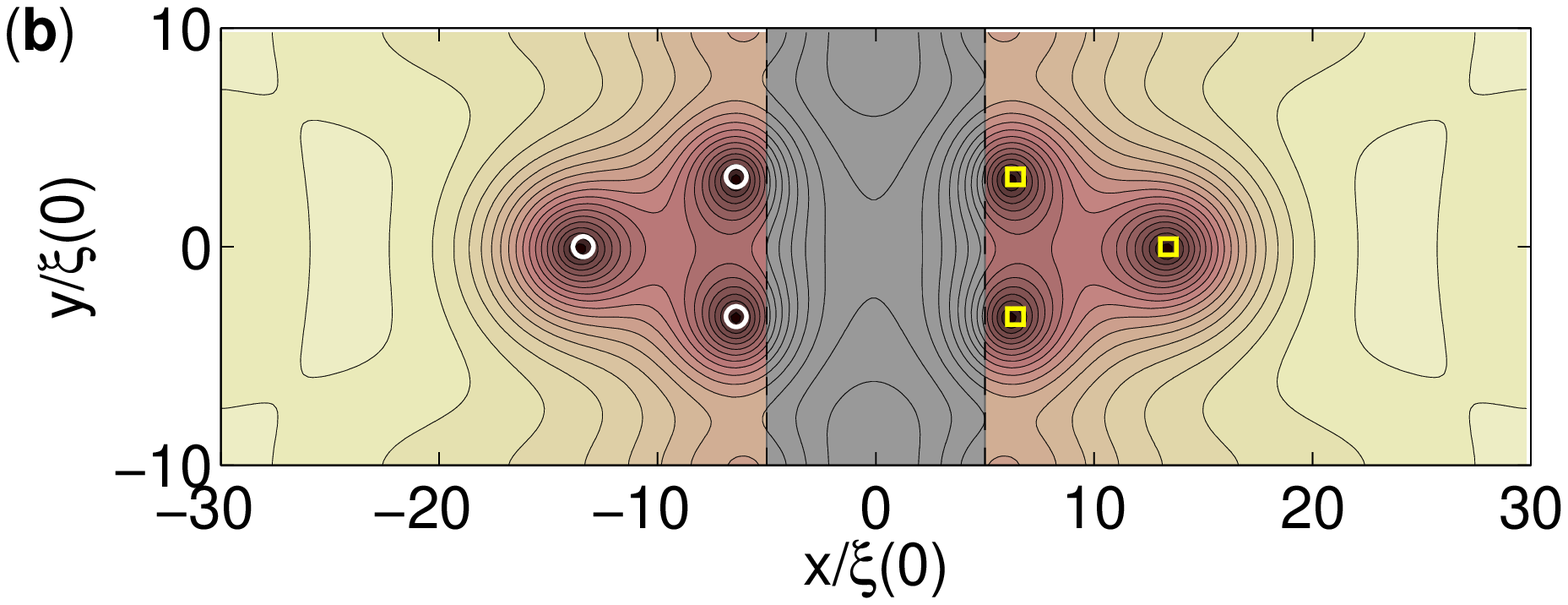}
    \includegraphics[width=6.9cm]{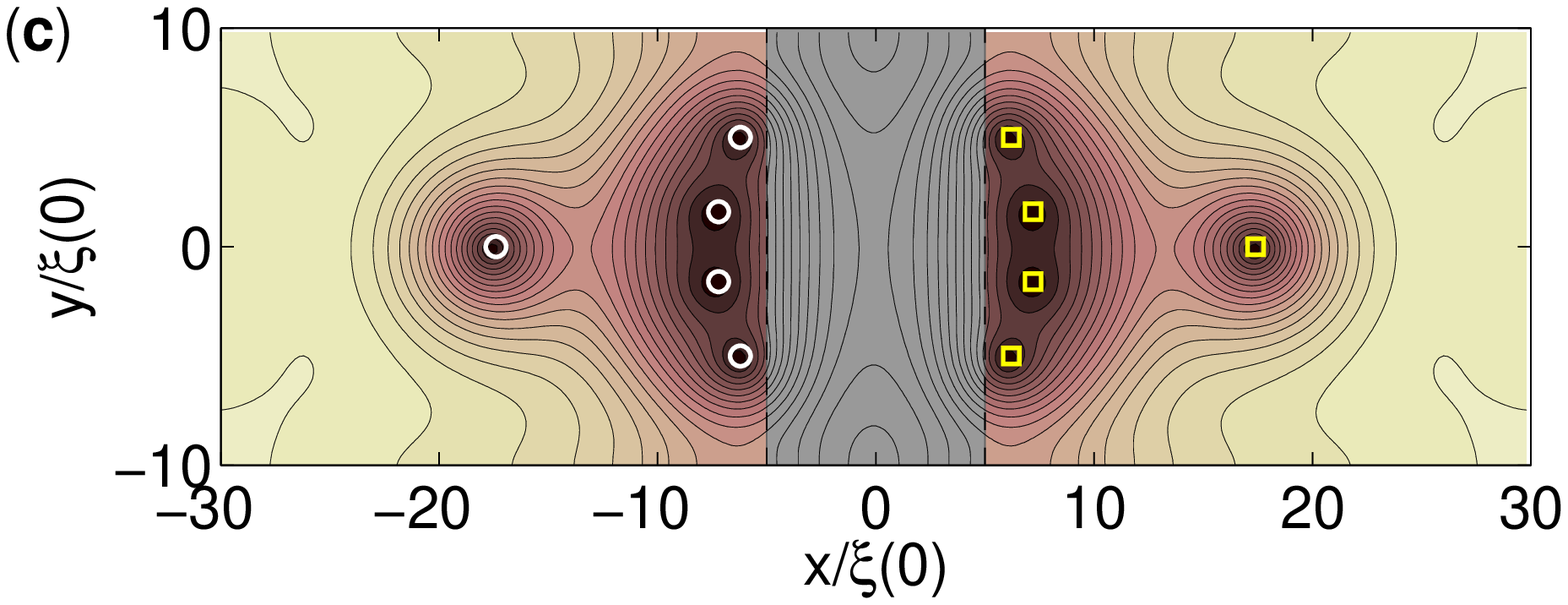}
    \includegraphics[width=6.9cm]{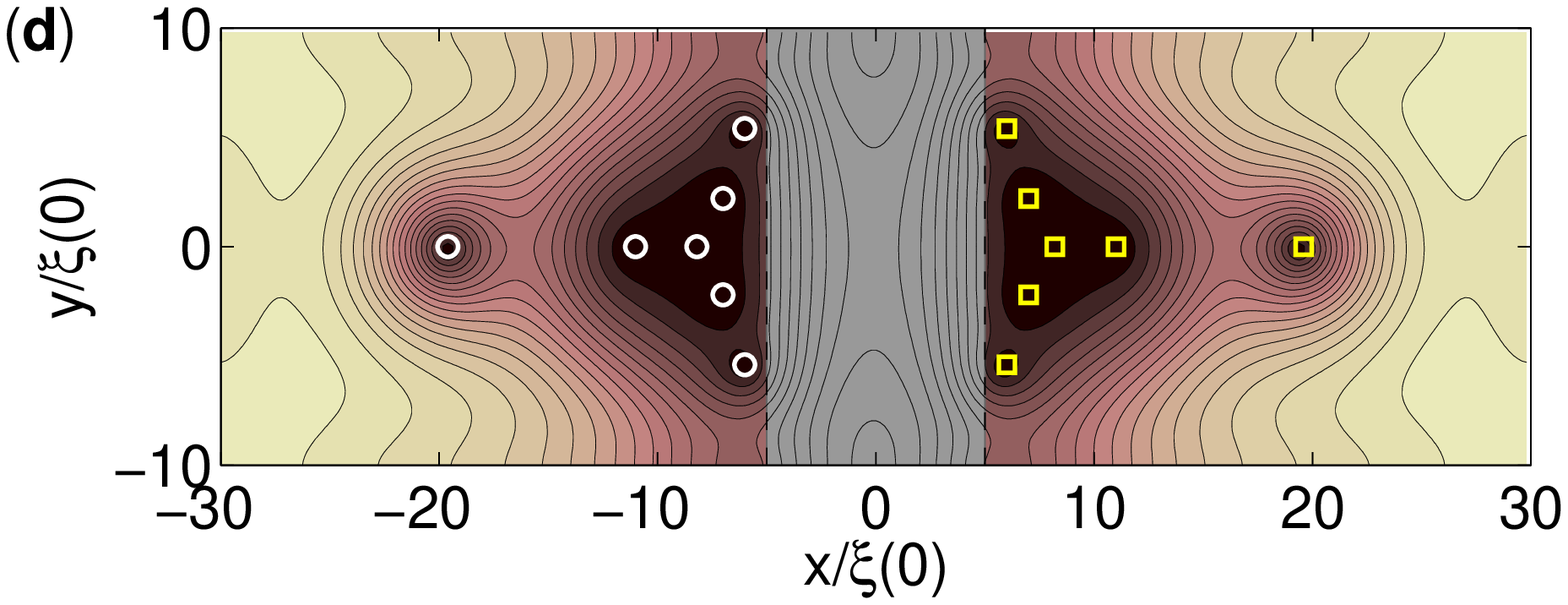}
    \end{center}
    \caption{(color online) Contour plots of $|\psi|$, showing equilibrium vortex-antivortex patterns in the thin
    superconducting strip (60$\,\xi(0)\times$20$\,\xi(0)$ in size), appearing  at $H=0$ and $T=0.9\,T_{c0}$ in the
    field of the current-carrying wire: $I_w=5$ mA (a), $I_w=8$ mA (b), $I_w=10$ mA (c), and $I_w=14$ mA (d).
    Lighter shades correspond to higher $|\Psi|$
    values, darker shades are the regions with suppressed
    superconductivity. The grey rectangle in the center of the strip
    is the projection of the current-carrying wire. The symbols depict the OP phase singularities
    (i.e., the cores of the vortices and antivortices).}
    \label{Fig-Equilibruim-0j9}
    \end{figure}

To facilitate further comparison between theory and experiment
(Section III), for our modeling we choose the parameters, typical
for mesoscopic Al-based superconductors: the coherence length
$\xi(0)=0.15\,\mu$m at $T=0$, length and width of the strip
$L_x=9\,\mu$m and $L_y=3\,\mu$m, width and thickness of the
current-carrying wire $L_w=1.5\,\mu$m and $D_w=0.05\,\mu$m, and
the separation between the strip and the wire $h=0.05\,\mu$m
(Fig.~\ref{Fig-System}).


\subsection{Mixed state in cross-film cryotron in equilibrium at $H=0$}


The stray field generated by the control wire is uniform across
the superconducting bridge (in the $y-$direction,
Fig.~\ref{Fig-System}b). However in the $x-$direction the field is
substantially inhomogeneous with the $b_z$--component reaching its
extreme values near the edges of the current-carrying wire. The
increase in the control current $I_w$ leads to a gradual
enhancement of the maximal $b_z$ value. This eventually results in
the local suppression of the energy barrier for vortex penetration
from the edges of the strip near the $|b_z|$ maxima. These regions
play the role of predefined gates for vortex entry.

    \begin{figure}[hb!]
    \begin{center}
    \includegraphics[width=8.5cm]{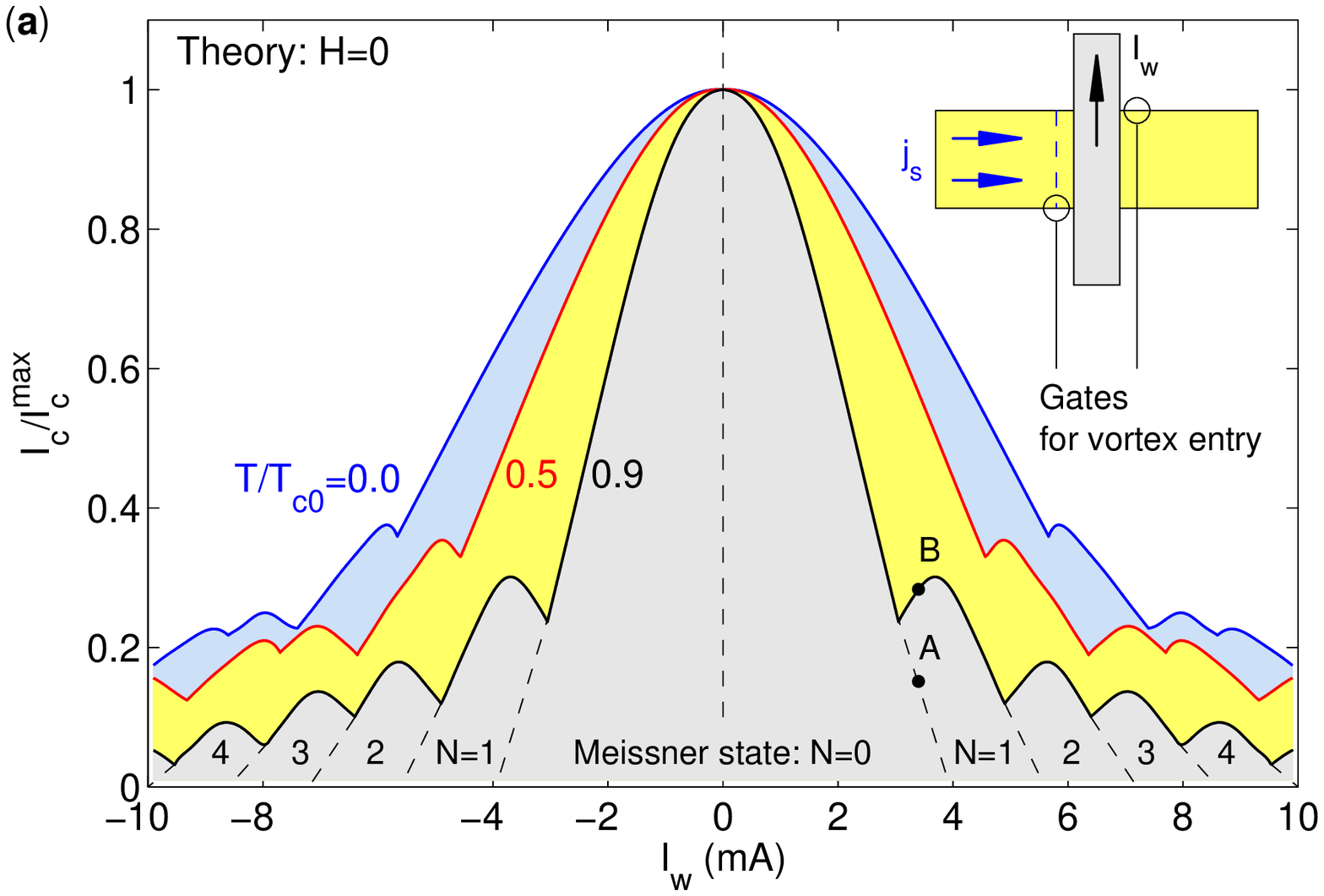}
    \includegraphics[width=8.5cm]{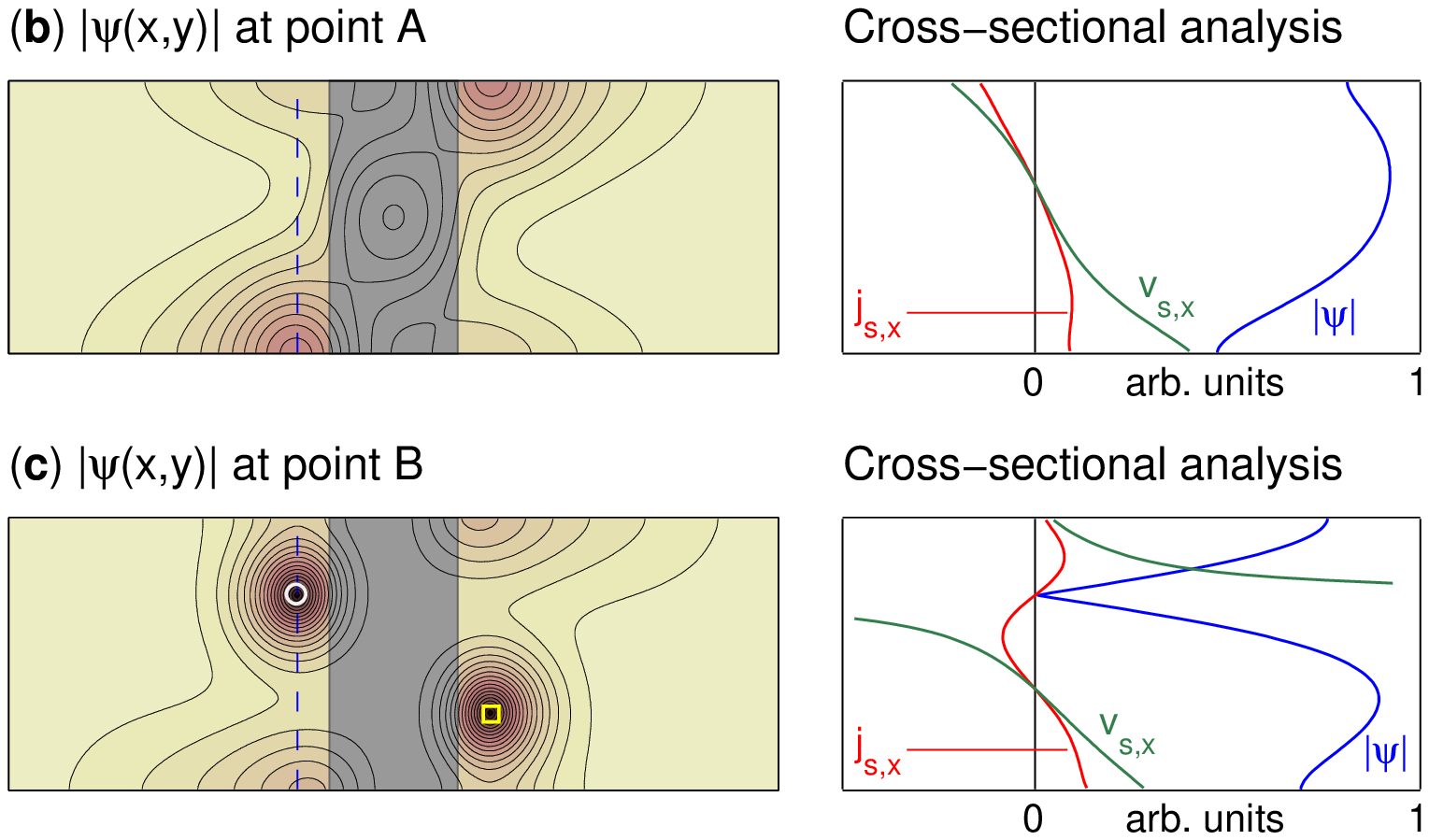}
    \end{center}
    \caption{(color online) (a) Critical current $I_c$ as a function of control current $I_w$ calculated
    at $H=0$ and $T/T_{c0}=0,\, 0.5,\,$ and 0.9. For each segment of the $I_c-I_w$
    diagram we indicate the number of the trapped vortices/antivortices
    $N$ in the strip. (b-c) Left panels show the stationary OP wave functions $|\psi|$
    calculated for $T/T_{c0}$ = 0.9 and $I_w$ = 3.4 mA for two different values of the injected current
    [points A and B in panel (a)]. The grey rectangle in the center of
    the strip is the projection of the current-carrying wire. The
    correspondent right panels show the profiles of the order
    parameter wave function $|\psi|$, the longitudinal $x-$components of the
    current $j_{s,x}$ and the supercurrent velocity
    $v_{s,x}\propto$ \mbox{$(\partial\theta/\partial x -2\pi A_x/\Phi_0$)} across the strip (i.e. along the dashed line).}
    \label{Fig-Theory-zeroH}
    \end{figure}

The stray field of the control wire acts as (i) a tunable source
of vortices and antivortices and, (ii) a vortex trap which
prevents vortex-antivortex pairs from their mutual annihilation
and the escape from the superconducting bridge. The latter follows
from the distribution of the screening superconducting currents
$j_s$ induced by the current-carrying wire (see
Fig.~\ref{Fig-Theory-zeroH}b). Indeed, the different signs of the
$x-$component of $j_s$ near the top and bottom edges of the strip
mean that the Lorentz force ${\bf f}_L=(\Phi_0/c)\,[{\bf
j}_s\times {\bf z}_0]$ acting on the vortex (antivortex) near
these edges will be oriented in such a way in order to push the
vortex (antivortex) to the inner part of the strip. In the absence
of an external magnetic field $H$ the appearance of a vortex
should be accompanied by the formation of a symmetrically
positioned antivortex and the number of vortex-antivortex pairs
$N$ increases as $|I_w|$ increases
(Fig.~\ref{Fig-Equilibruim-0j9}). Since the $b_z-$field roughly
decays inversely proportional to the distance from the wire, the
vortices and antivortices are also distributed non-uniformly: the
closer to the wire, the smaller the distance between two vortices
(antivortices) is.

\subsection{Critical current of cross-film cryotron at $H=0$}

Depending on the magnitude of the bias current $I_{ext}$, injected
into the superconducting strip, there are two distinct regimes.
After applying a small bias current the disturbed vortex and
antivortex tend to relax into a new stationary configuration. This
state with motionless vortex-antivortex pairs is characterized by
the absence of an electrical field inside the superconductor
(except for tiny regions near the boundaries where the current
injection takes place) and by a zero voltage drop. If $I_{ext}$
exceeds the critical current $I_c$, the stability of the
vortex-antivortex ensemble breaks down: vortex and antivortex
periodically in time enter and exit into/from the superconductor.
This means that the sample has switched to a dissipative regime.
The calculated dependencies $I_c(I_w)$ at $H=0$ for different
temperatures are shown in Fig.~\ref{Fig-Theory-zeroH}a. We see
that $I_c$ decays with oscillations as $I_w$ sweeps and the $I_c$
oscillations become more pronounced at high temperatures.

    \begin{figure}[ht!]
    \begin{center}
    \includegraphics[width=8.5cm]{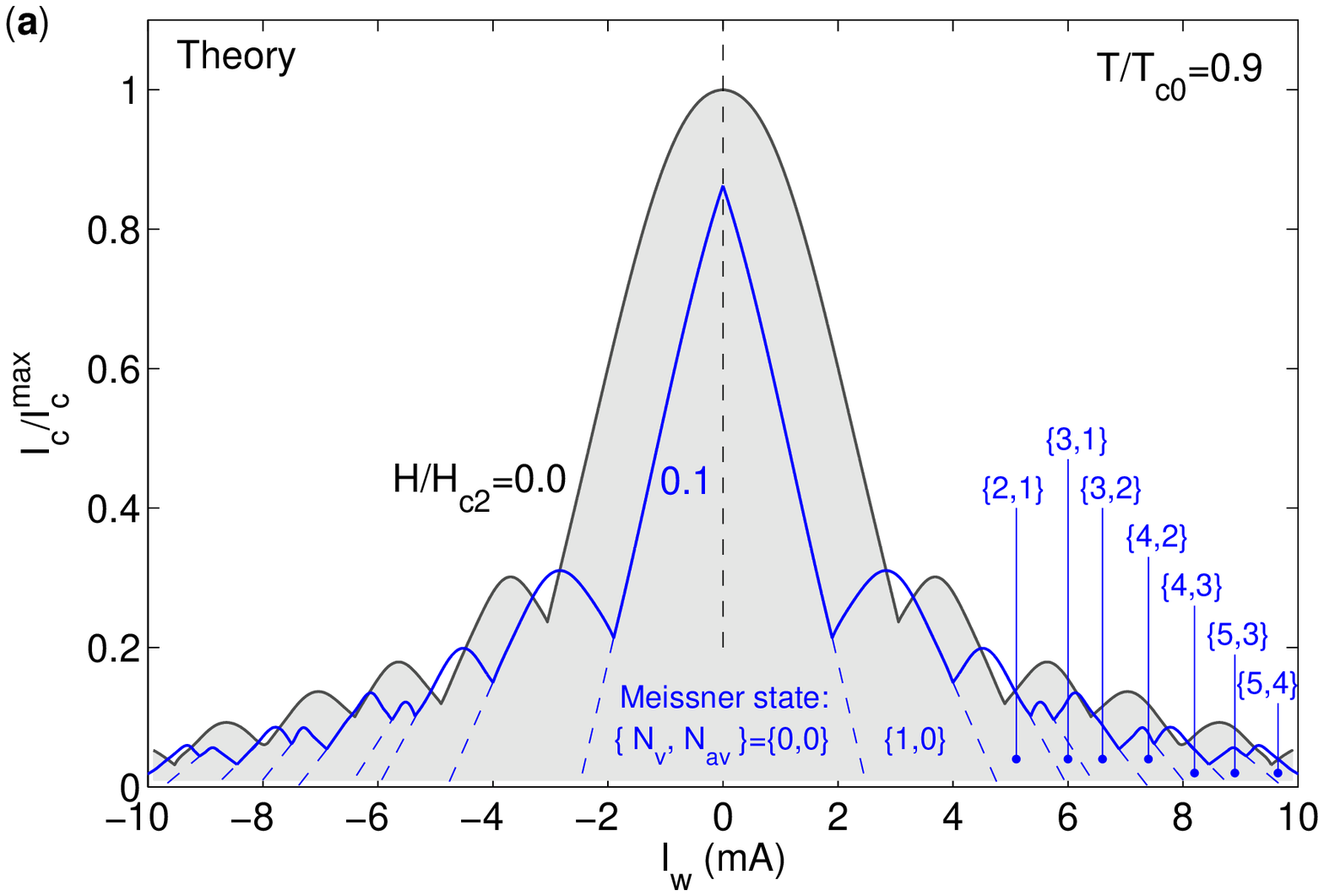}
    \includegraphics[width=4.2cm]{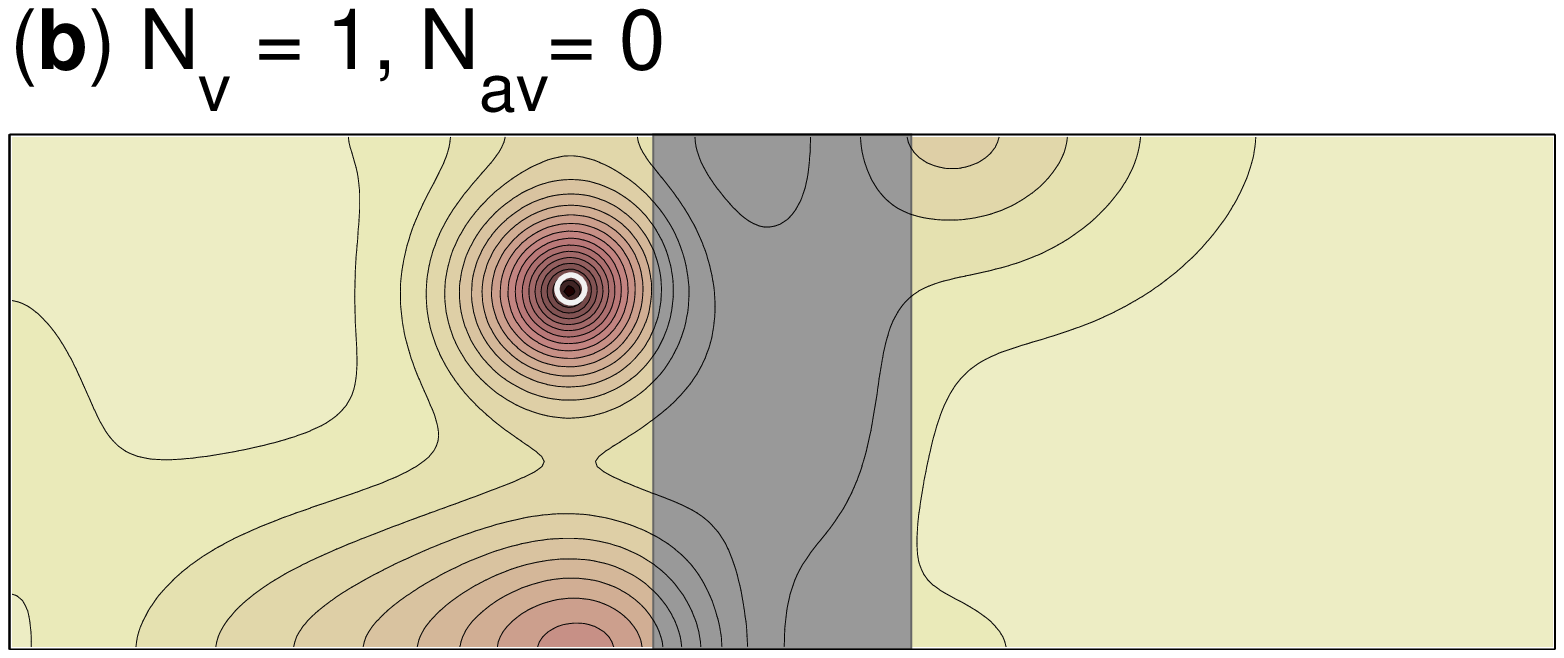}
    \includegraphics[width=4.2cm]{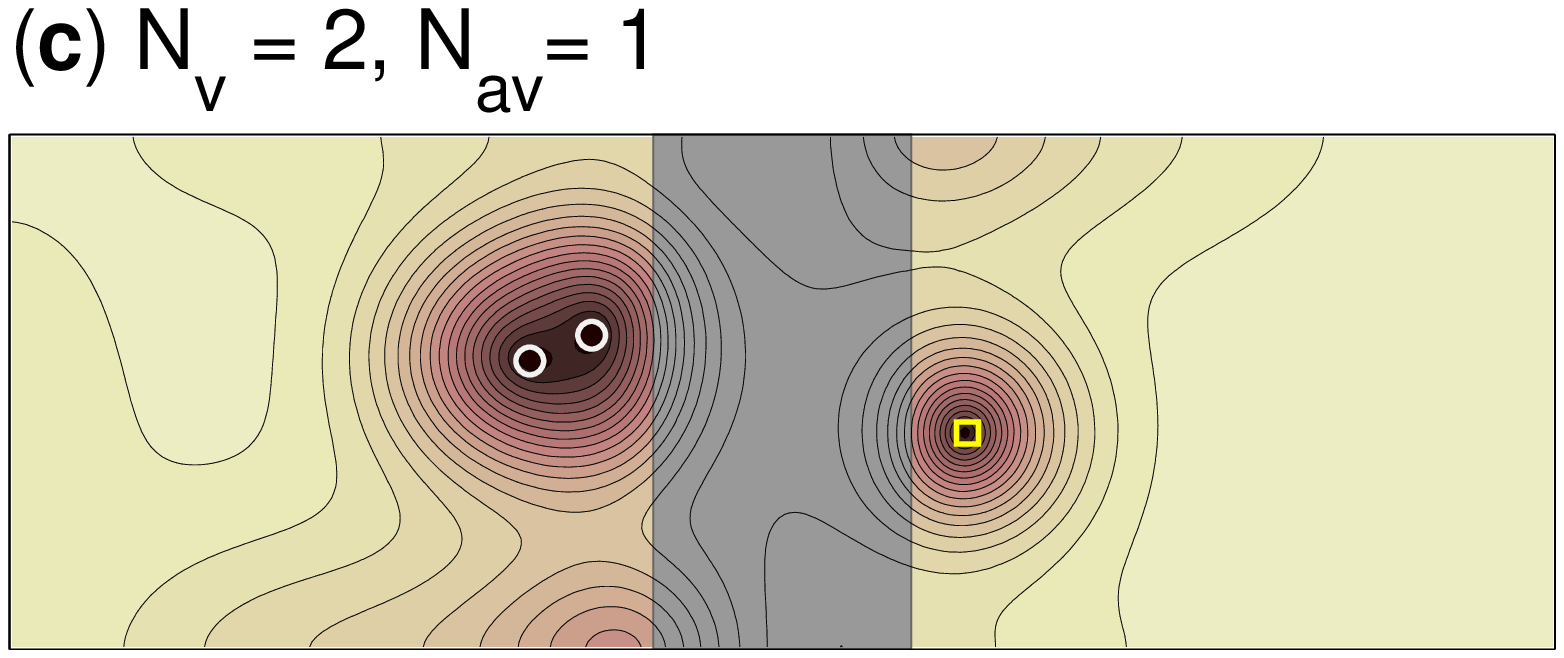}
    \includegraphics[width=4.2cm]{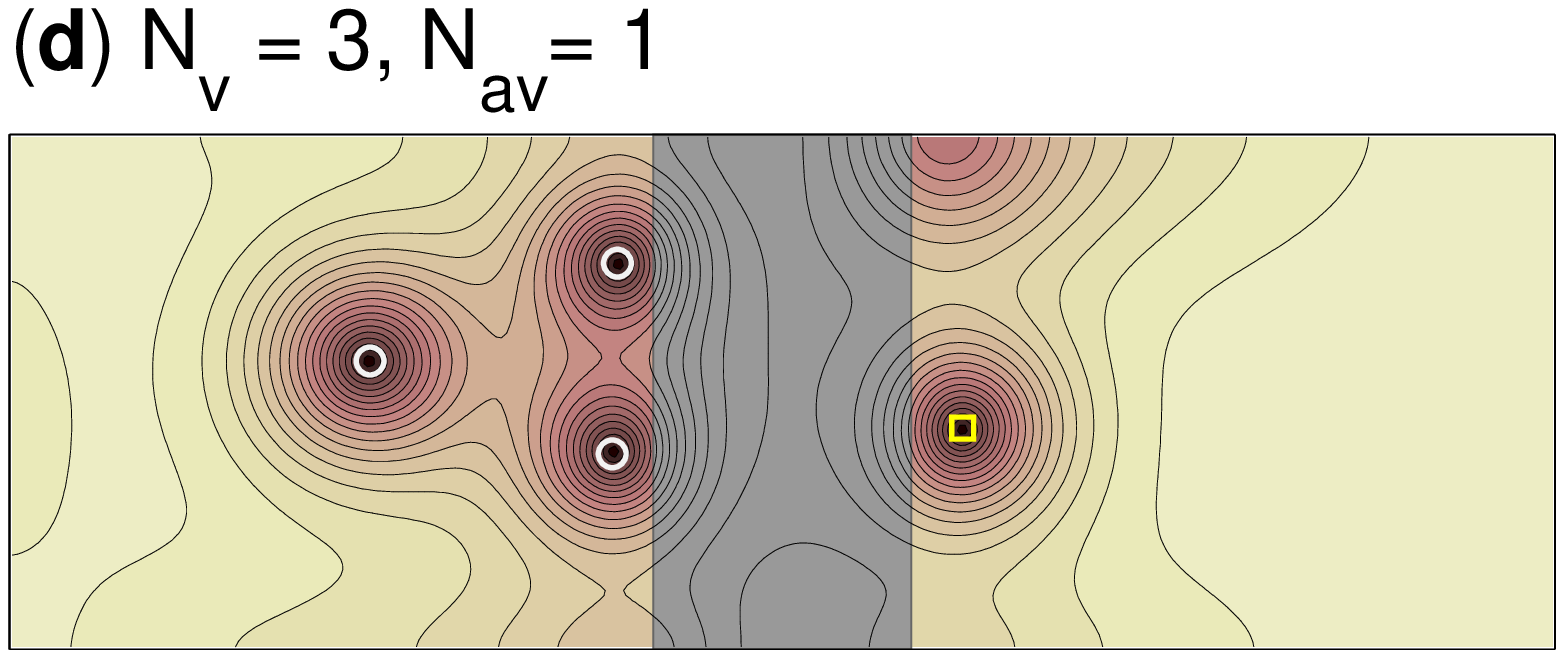}
    \includegraphics[width=4.2cm]{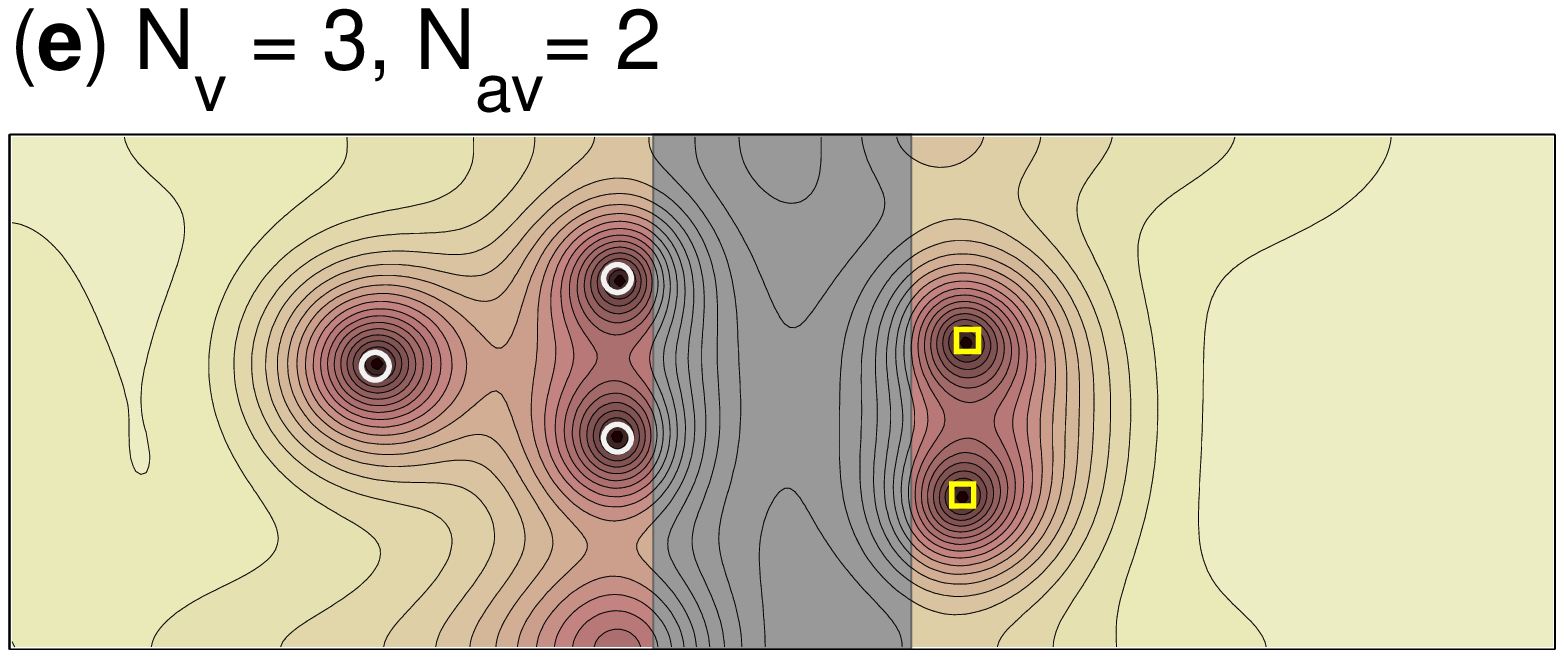}
    \end{center}
    \caption{(a) The calculated dependencies $I_c(I_w)$ for $T/T_{c0}=0.9$ and $H=0$ (black line) and $H/H_{c2}=0.1$ (blue line).
    For each segment of the $I_c-I_w$ diagram we indicate the number of vortices $N_v$ and
    antivortices $N_{av}$, trapped in the superconducting strip. (b--e)
    Examples of the stationary OP patterns
    at $T/T_{c0}=0.9$ and $H/H_{c2}=0.1$ and at bias current close to the corresponding critical values.}
    \label{Fig-Theory-NoNzeroH}
    \end{figure}

We can explain the appearance of the $I_c(I_w)$ oscillations as
follows. If $H=0$ and $I_w=0$, superconductivity will survive
until the injected current density exceeds the depairing limit.
For nonzero $I_w$ the transition from the non-dissipative state to
the resistive regime occurs via permanent formation of
vortex-antivortex pairs at the opposite edges of the
superconducting bridge and their motion across the bridge (see
inset in Fig.~\ref{Fig-Theory-zeroH}). Due to the superposition of
the injected current and the currents, induced by the wire, the
limit for the vortex-antivortex pair generation can be reached at
smaller values of the bias current; therefore  $I_c$ monotonously
decreases as $I_w$ increases. A further increase in $I_w$ leads to
(i) a decrease in the threshold value of the bias current required
for the creation of the first vortex, and (ii) an enhancement of
the trapping potential for vortices. Both consequences make it
possible to stabilize the vortex-antivortex pair in the presence
of the transport current (Fig.~\ref{Fig-Theory-zeroH}c). The
self-currents generated by the vortex and antivortex partly
compensate the superflow conditioned by the bias current near the
regions with suppressed OP wave function. Therefore these ``gates"
will be {\it closed} and the formation of new vortices and
antivortices will be {\it impeded} unless the maximal value for
supercurrent velocity ${\bf v}_s$, which is proportional to
$\nabla\theta-2\pi{\bf A}/\Phi_0$, again reaches its critical
value at the strip edges. This means that after the first
vortex-antivortex pair is trapped, the critical current will be
larger than that in the Meissner state. Then this process is
repeated periodically as $I_w$ increases, resulting in periodic
variations in $I_c(I_w)$. Thus, the discrete change in the number
of the pinned vortices and antivortices in the superconducting
strip of a finite width can be identified by considering the
position of the cusps on the $I_c(I_w)$ curve. In contrast to the
$I_c$ oscillations observed in mesoscopic superconducting squares
in a uniform magnetic field,\cite{Vololazov-PRB-2005} the
nonuniform magnetic field gives us the possibility to see the
quantization effects in a {\em long} superconducting strip, due to
the confinement potential along the strip produced by the control
wire.

The absolute value of the magnetic flux $\Phi_{1/2}$ piercing a
half of the superconducting strip,
    \begin{eqnarray}
    \Phi_{1/2}=\int\limits_{0}^{L_x/2} \int\limits_{-L_y/2}^{L_y/2}
    |b_z(x,y)|\,dxdy,
    \end{eqnarray}
is a linear function of $I_w$. In a certain sense the oscillations
of the critical current of the cryotron as $I_w$ (or $\Phi_{1/2}$)
varies, are similar to the standard Fraunhofer pattern describing
the dependence of the critical current of a Josephson junction on
the total magnetic flux $\Phi$ through the junction
area.\cite{BaronePaterno-82,Abrikosov-88} The built-in field of
the wire guarantees reversible entrance of vortices and
antivortices into the superconducting strip as $I_w$ increases,
and the tunable modification of the distribution of the OP phase
in the restricted part of the strip by the trapped vortices and
antivortices. This area with partly suppressed superconductivity
near the control wire acts like an effective {\em weak link},
since the vortex dynamics in this area determines the flow of the
bias current and the resistance of the entire strip.

\subsection{Mixed state and critical current of cross-film cryotron at $H\neq
0$}

The external magnetic field $H$, applied perpendicularly to the
sample's plane, breaks the symmetry between vortex and antivortex
since the total magnetic flux piercing the sample is not zero
anymore. As a result, the entrance of a new vortex is not
generally accompanied by the entrance of an antivortex: the
numbers of trapped vortices and antivortices, $N_v$ and $N_{av}$,
may differ in contrast to the previously considered case $H=0$.

The results of the calculations for $T/T_{c0}=0.9$ and at
$H/H_{c2}=0$ and $H/H_{c2}=0.1$ are compared in
Fig.~\ref{Fig-Theory-NoNzeroH}a. It is clear that the oscillating
behavior in the $I_c$ vs. $I_w$ remains even in the presence of
the applied magnetic field. Since the constructive superposition
of (i) the injected current, (ii) the currents induced by the wire
and (iii) the currents induced by the external field, enhances the
local flow of superconducting condensate, the conditions for
vortex entry will be fulfilled at smaller $I_w$ values. This
results in a shift of all the cusps on the $I_c(I_w)$ dependence
towards to zero. The non-equivalence between vortices and
antivortices allows us to get new stable states of reduced
symmetry (Fig.~\ref{Fig-Theory-NoNzeroH}b-e) which are forbidden
for the symmetrical sample at $H=0$. The transition between such
states sweeping $I_w$ are accompanied by additional cusps, which
leads to more complex oscillatory properties of the $I_c$ on $I_w$
dependence.

\section{Experiment and discussion}

\begin{figure}[ht!]
    \centering
    \includegraphics[width=7.3cm]{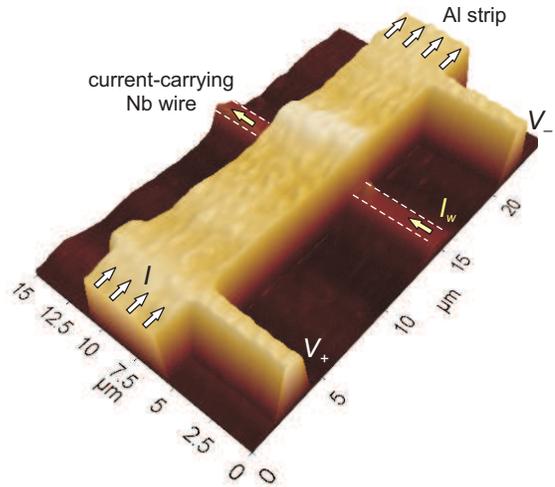}
    \caption{(color online) AFM image of the sample: the top (light) element is the superconducting Al
    strip with two perpendicular contacts for the measurement of the voltage drop, the bottom element oriented perpendicular to the strip is
    the current-carrying Nb wire.
    \label{Fig-AFM-Experiment}}
    \end{figure}

To study the transport properties of the mesoscopic cross-film
cryotron, we fabricated the hybrid structures consisting of a
$4~\mu$m wide and 120 nm thick superconducting Al strip on top of
a 1.5 $\mu$m wide and 50 nm thick current-carrying Nb wire. The Nb
thin film was deposited via dc magnetron sputtering at a rate of
2.5 nm/s on an oxidized Si substrate at room temperature under the
pressure of Ar plasma of 6$\times 10^{-3}$~mbar.  The Nb wire was
fabricated by e-beam lithography in combination with argon ion
milling. The Al film was grown in a molecular-beam epitaxy
apparatus at a rate of 0.2~nm/s at room temperature and nominal
pressure of $1\times 10^{-8}$~mbar. The Al strip was patterned by
standard e-beam lithography, subsequent evaporation and lift-off
technique. A 120 nm thick Ge layer separates the two metals from
each other to avoid electrical contact between them. We tested
more than 30 identical hybrid samples prepared at different times
but under similar conditions. At least 10 samples have no problems
with electrical leakage, since the applying a voltage between Al
strip and Nb wire leads to an absence of detectable current. It
gives us a lower limit on the contact resistance of the order of
5~M$\Omega$. An Atomic Force Microscopy (AFM) image of the sample
is shown in Fig.~\ref{Fig-AFM-Experiment}.

\begin{figure}[hb!]
    \centering
    \includegraphics[height=4.5cm]{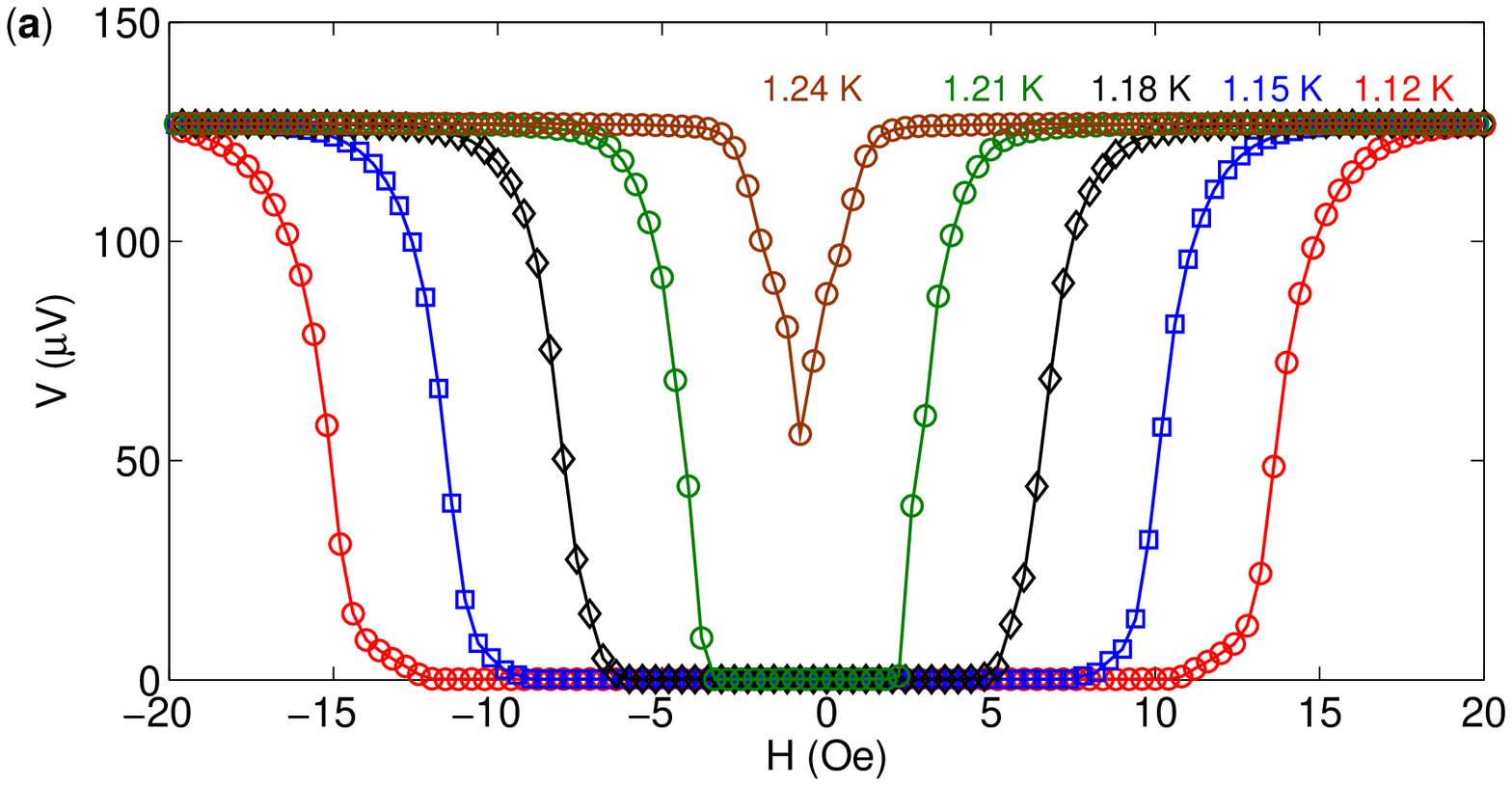}
    \includegraphics[height=4.5cm]{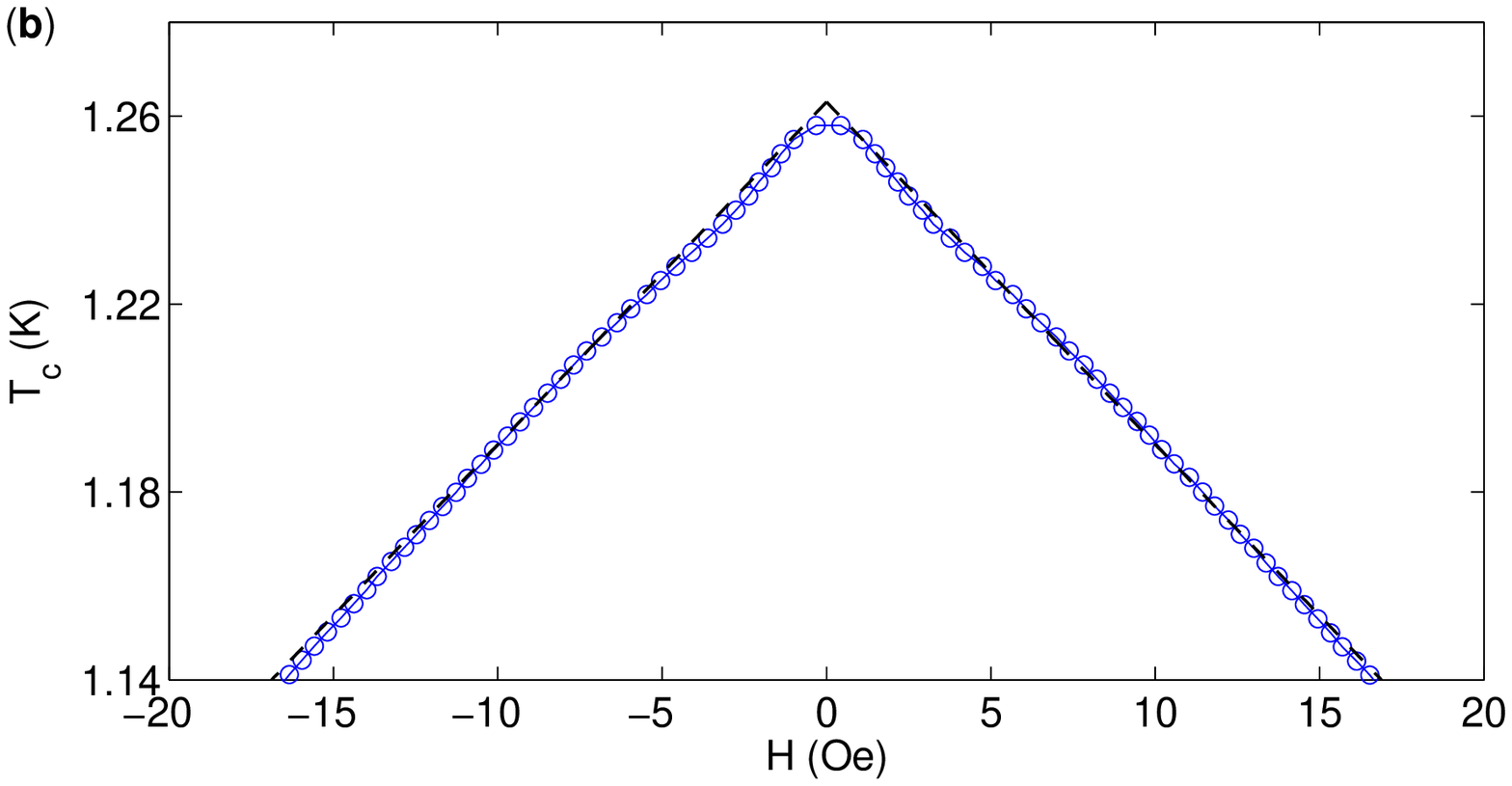}
    \caption{(color online) (a) Typical dependence of the voltage drop,
    $V$, as a function of $H$ at $I_w=0$ for different
    temperatures. The bias dc current $I$ is equal to 50~$\mu$A,
    the normal state resistance $R_n\simeq 0.45~\Omega$.
    (b) The phase boundary $T_c(H)$ extracted
    from the magnetoresistive measurements at $I_w=0$ according
    to the criterion $R(H,T)=0.99\,R_n$. The dashed line is the
    linear approximation
    $(1-T/T_{c0}) = 0.59\, |H|/H_{c2}^{(0)}$, where $H_{c2}^{(0)}\simeq
    104~$Oe, $T_{c0}=1.265~$K.
    \label{Fig-PB-Experiment}}
    \end{figure}

Considering the dependence of the sample resistance $R$ on $H$ and
its displacement upon varying $T$, one can compose the phase
transition line $T_c(H)$ shown in Fig.~\ref{Fig-PB-Experiment}b.
By identifying this line with the temperature dependence of the
critical field of surface superconductivity
$H_{c3}=1.69\,H_{c2}^{(0)}\,(1-T/T_{c0})$, we estimated
$T_{c0}\simeq 1.265~$K, the upper critical field
$H_{c2}^{(0)}\simeq 104$~Oe and the coherence length
$\xi(0)=\big(\Phi_0/2\pi H_{c2}^{(0)}\big)^{1/2}\simeq 175~$nm
extrapolated to $T=0$. The mean free path $\ell\simeq 35$ nm can
be obtained from the value of the normal resistivity $\rho_n$ at
low temperatures according to the formula:\cite{Fickett-71}
$\rho_n\,\ell \simeq 6\times 10^{-16}~\Omega\cdot$m$^2$.  The
$\ell\sim 35$~nm value gives us the London penetration depth and
the effective magnetic field penetration depth $\lambda_L(0)\simeq
175$ nm and $\Lambda(0)=\lambda^2_L(0)/d\simeq 250\,$nm
respectively in dirty limit at $T=0$. All obtained values are in
accordance with previous studies on thin film Al
structures.\cite{Gillijns-07}

    \begin{figure}[ht!]
    \centering
    \includegraphics[width=8.3cm]{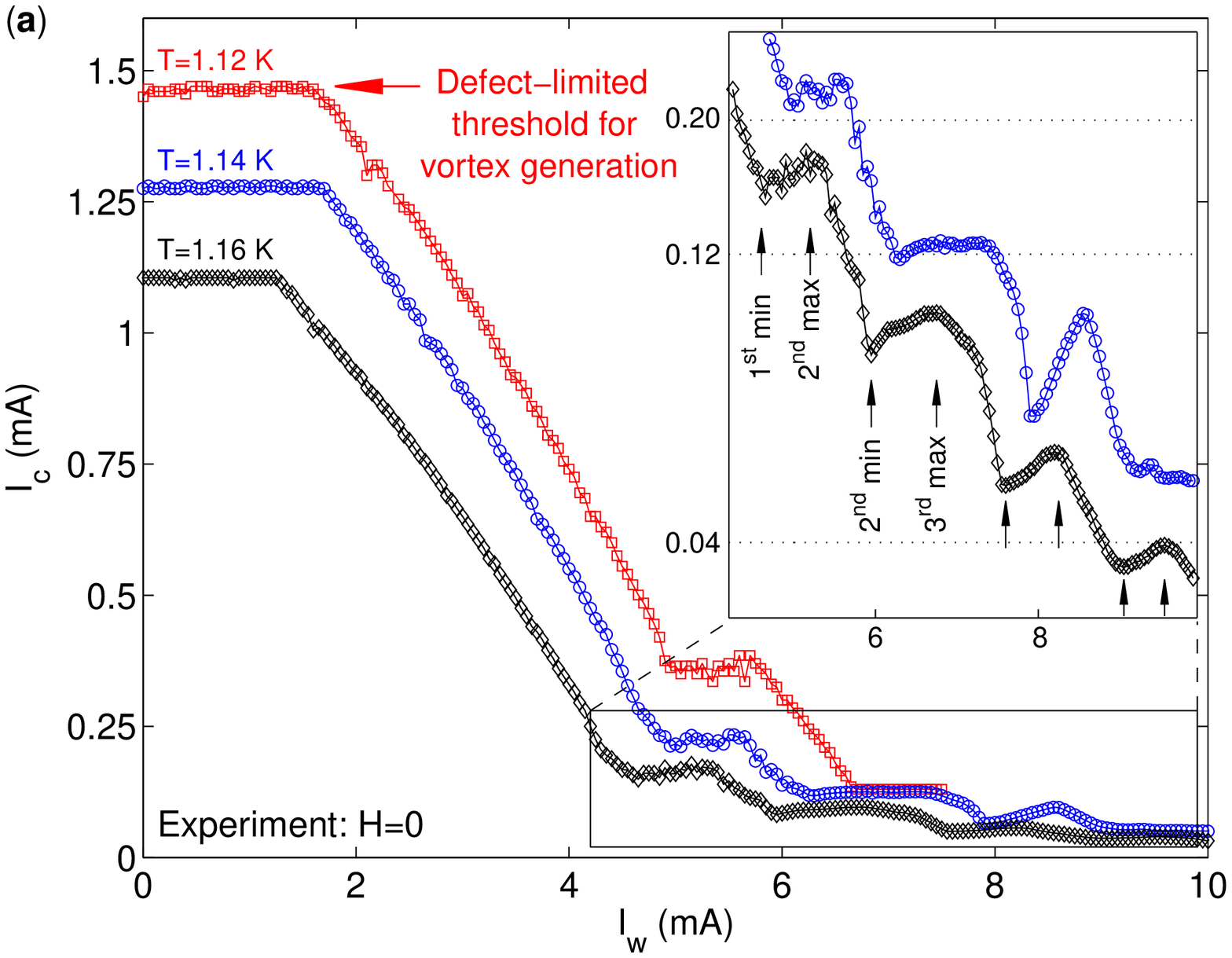}\\
    \includegraphics[width=8.3cm]{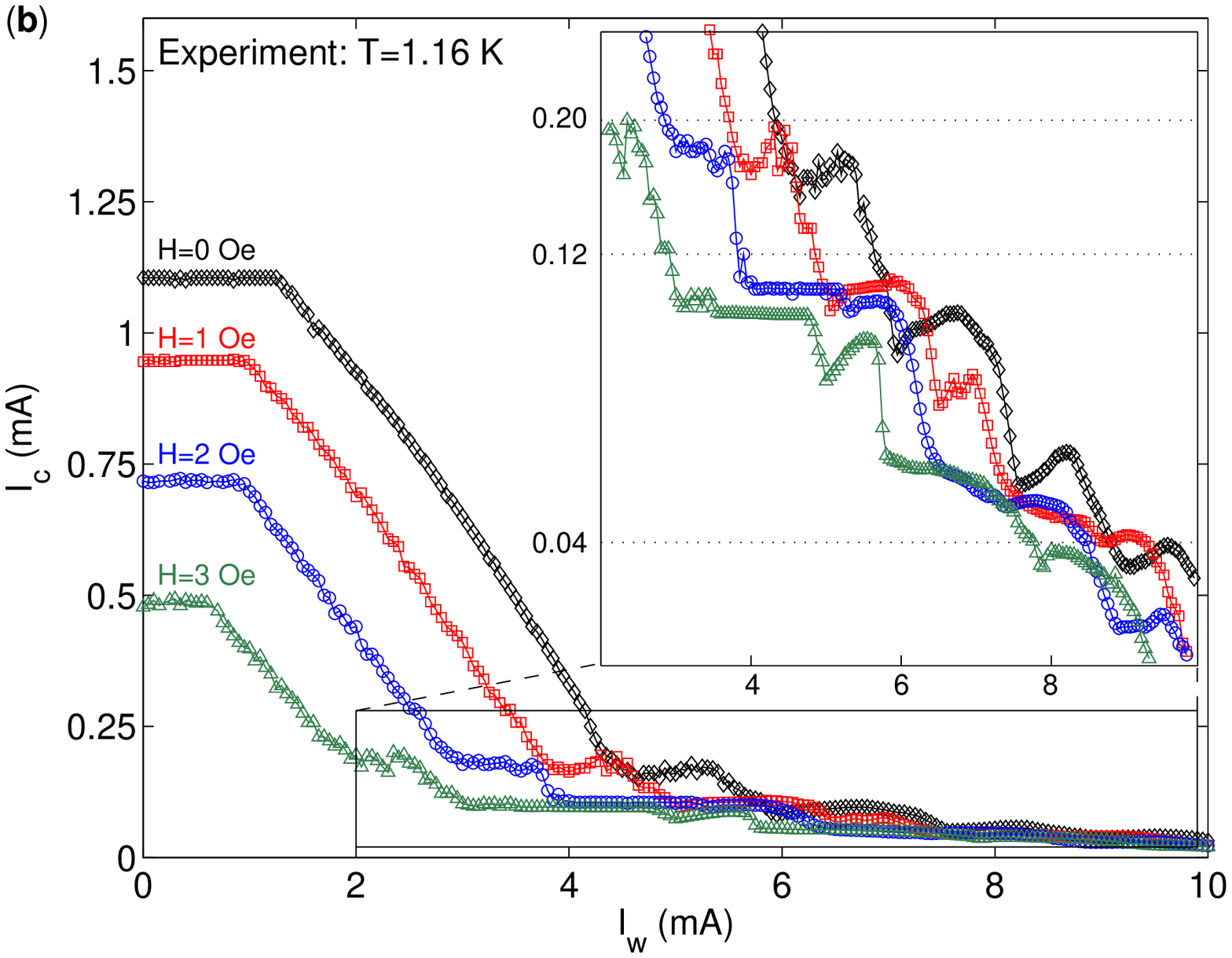}\\
    \caption{(color online) (a) Measured dependence
    $I_c$ on $I_w$ at $H=0$ for different temperatures.
    (c) Measured dependence
    $I_c$ on $I_w$ at $T=1.16\,$K for different $H$ values. Both
    insets in the panels (b) and (c) show zoomed parts of the $I_c(I_w)$ curves (within the
    box) in semi-logarithmic scale. The arrows in the inset (a) show the positions of the extremal $I_c$ values for $T=1.16~$K and $H=0$,
    which are compared with theory in Fig.~\ref{Fig-Comparison}.
    The confidence intervals are smaller than the size of symbols,
    and therefore error bars cannot be shown in this scale.
    \label{Fig-CritCurrent-Experiment}}
    \end{figure}


Figure \ref{Fig-CritCurrent-Experiment}a shows the $I_c$ of the Al
strip as a function of the current $I_w$ in the control Nb wire,
measured at $H=0$ and different temperatures close to $T_{c0}$.
Experimentally the critical current was determined from the
isothermal current $(I)$ -- voltage $(V)$ dependencies for
different voltage criteria (from 0.1~$\mu V$ up to 10~$\mu V$
between the two voltage contacts at a distance 20 micron) for the
measured voltage drop.\cite{Campbell-Evetts} For our measurements
the accuracy $\delta I_c$ of the determination of the critical
current was 3~$\mu$A. This means that within the interval
2$\,\delta I_c$ we observed a sharp transition from the
superconducting state (typical noise level of the order of
$5\times 10^{-8}$ V) to the resistive state. The plateau in
$I_c(I_w)$ might be associated with the influence of sample
imperfections, which facilitate the formation of vortices near
defects even in zero magnetic field at current densities smaller
than the theoretical limit. In accordance with our simulations the
monotonously decreasing part of $I_c(I_w)$ has to be attributed to
the transition from the Meissner state to the resistive state with
a single moving vortex-antivortex pair. The next stages correspond
to the stabilization of the first and all subsequent
vortex-antivortex pairs in the sample. The switching between these
states at varying $I_w$ corresponds to the local $I_c$ minima. We
clearly observed the reproducible oscillations of the critical
current for all measured samples.

The effect of the external field is illustrated by
Fig.~\ref{Fig-CritCurrent-Experiment}b. We observed the
anticipated shift of the monotonously decreasing part in
$I_c(I_w)$ towards smaller $I_w$ values, conditioned by the
constructive superposition of the injected current, the currents
induced by the wire and by the external field. This ensures that
the conditions for the first vortex entry will be fulfilled at
smaller $I_w$ values. It is interesting to note that the shape of
the oscillating $I_c(I_w)$ curves  becomes more complicated as $H$
increases. We suppose that the external-field-induced modification
of $I_c(I_w)$ reflects the formation of exotic non-symmetrical
vortex states and their depinning under the action of the bias
current.

    \begin{figure}[ht!]
    \begin{center}
    \includegraphics[width=8.5cm]{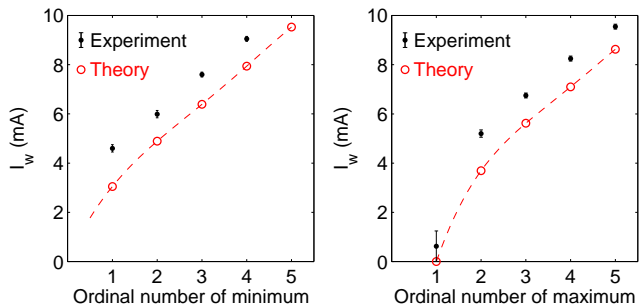}
    \end{center}
    \caption{(color online) Positions of the $I_c$ minima (left panel) and the $I_c$
    maxima (right panel), calculated at $H=0$ and $T/T_{c0}=0.9$ (red circles)
    and measured at $T=1.16$~K ($T/T_{c0}\simeq 0.91$, black dots). Dashed red lines are guide to the eyes. The
    positions of the maxima and minima determined experimentally are also marked by vertical
    arrows in the inset of Fig.~\ref{Fig-CritCurrent-Experiment}a.
    \label{Fig-Comparison}}
    \end{figure}

To compare theory and experiment, we plot the positions of the
minima and maxima in the $I_c(I_w)$ dependence at $H=0$
(Fig.~\ref{Fig-Comparison}). It is easy to see that our simple
model describes both the general oscillating behavior for the
$I_c(I_w)$ dependence for real sample and the observed period
$\Delta I_w$ quite well. The fact that all theoretical points in
Fig.~\ref{Fig-Comparison} lie below the experimental data, may be
caused, e.g., by a nonuniform current distribution in the control
wire, or by a small difference between the parameters of the model
problem and the dimensions of the tested samples. Nevertheless
using the estimated period $\Delta I_w\simeq 1.45$ mA, one can
calculate the change in the magnetic flux $\Delta \Phi_{1/2}$
through the half of the superconducting strip and corresponding to
the appearance of the vortex-antivortex pair:
$\Delta\Phi_{1/2}\simeq 1.1\Phi_0$.


In this paper we have reported on the first results obtained for
the prepared S/Em samples. We note that the oscillatory behavior
of the critical current as a function of $I_w$ was found to be
reproducible for all measured samples, which had identical
geometry and size as the presented device. However the issues
concerning the influence of the dimensions of the superconducting
strip and the current-carrying wire on the $I_c$ oscillations
remain unaddressed.


\section{Conclusion}

We studied the transport properties of a hybrid system consisting
of a superconducting strip and a current-carrying wire, oriented
perpendicular to the strip. The stray field, generated by the
control wire, makes it possible to tune the transport properties
of the long superconducting strip locally by means of creation and
pinning of vortices and antivortices. This area, where vortices
and antivortices become confined, plays the role of a
``bottleneck" for the transport current, since the motion of
vortices and antivortices is the main source of the dissipation.
Thus, the peculiar configurations of vortex-antivortex pairs
determine the critical current $I_c$ of the S strip. The
transition between different stable vortex patterns upon varying
$I_w$ results in an oscillatory dependence of $I_c$ on $I_w$. The
observed effect, which looks similar to the Fraunhofer oscillating
pattern for conventional Josephson junctions, can be potentially
interesting for the design of superconducting interferometer
devices based on S/F or S/Em hybrid structures.

This work was supported by the Methusalem Funding of the Flemish
Government, the NES -- ESF program, the Belgian IAP, the
Carl-Zeiss-Stiftung and the Deutsche Forschungsgemeinschaft (DFG)
via the SFB/TRR 21, the Russian Fund for Basic Research, RAS under
the Program ``Quantum physics of condensed matter" and Federal
Target Program ``Scientific and educational personnel of
innovative Russia in 2009--2013". A.V.S, W.G. and J.V.d.V.
acknowledge support from F.W.O.

%
%

\end{document}